\documentstyle[12pt]{article}
\vspace{1.8cm} \oddsidemargin 0pt \evensidemargin 0pt  \topmargin
-1.5cm \footheight 26pt \footskip 10mm

\begin{document}
\begin{center}
\baselineskip=18pt

{\bf Quantum wave equation of non-conservative system}

\vspace{1cm}{Xiang-Yao Wu$^{a}$ \footnote{E-mail:wuxy2066@163.com
}, Bai-Jun Zhang$^{a}$, Hai-Bo Li$^{a}$ \\ Xiao-Jing Liu$^{a}$,
Jing-Wu Li$^{a}$ and Yi-Qing Guo$^{b}$}

\vskip 10pt \noindent{\footnotesize a. \textit{Institute of
Physics,
Jilin Normal University, Siping 136000, China}\\
\footnotesize b. Institute of High Energy Physics, P. O. Box
918(3), Beijing 100049, China}

\end{center}
\date{}
\renewcommand{\thesection}{Sec. \Roman{section}} \topmargin 10pt
\renewcommand{\thesubsection}{ \arabic{subsection}} \topmargin 10pt
{\vskip 5mm
\begin {minipage}{140mm}
\centerline {\bf Abstract} \vskip 8pt
\par
\indent\\

\hspace{0.3in}It is well known that Schr\"{o}dinger's equation is
only suitable for the particle in conservative force field. In
atomic and molecular field, a particle can suffer the action of
non-conservative force. In this paper, a new quantum wave equation
is proposed, which can describe the particle in
non-conservative force field. We think the new quantum wave equation can be used in many fields.


\end {minipage}

\newpage
\section * {1. Introduction }

\hspace{0.3in}It is well known that quantum mechanics (QM)
acquired its final formulation in 1925-1926 through fundamental
papers of Schr\"{o}dinger and Heisenberg. Originally these papers
appeared as two independent views of the structure of quantum
mechanics, but in 1927 Schr\"{o}dinger established their
equivalence, and since then one or the other of the papers
mentioned have been used to analyze quantum mechanical systems,
depending on which method gave the most convenient way of solving
the problem. In the 1940's Richard Feynman, and later many others,
derived a propagator for quantum mechanical problems through a
path integration procedure[1-3]. In contrast with the Hamiltonian
emphasis in the original formulation of quantum mechanics,
Feynman's approach could be referred to as Lagrangian and it
emphasized the propagator $K(x, t; x', t')$ which takes the wave
function $\psi(x', t')$ at the point $x'$ and time $t'$ to the
point $x$ at time $t$. While this propagator could be derived by
the standard methods of quantum mechanics. Feynman invented a
procedure by summing all time dependent paths connecting points
$x, x'$ and this became an alternative formulation of quantum
mechanics whose results coincided with the older version when all
of them where applicable.

Quantum Mechanics has become one of the most important foundations
of physics, and achieved great success, physicists had begun to
consider the possibility to generalize the traditional framework
of it[4]. Up to now,  this kind of attempts have never
stopped[5-8]. As we known, Schr\"{o}dinger's equation is only
suitable for the particle lying in conservative force field. For a
non-conservative system, a number of other methods have been
proposed to study it. In Ref. [9], Riewe extend Lagrangian and
Hamiltonian mechanics to include derivatives of fractional order.
The fractional derivatives lead directly to equations of motion
with nonconservative classical forces such as friction. In Ref.
[10], Riewe continues the development of fractional-derivative
mechanics by deriving a modified Hamilton¡¯s principle. By using
fractional derivatives, it is possible to construct a complete
mechanical description of nonconservative systems, including
Lagrangian and Hamiltonian mechanics, canonical transformations,
Hamilton-Jacobi theory, and quantum wave mechanics. The method is
illustrated with a frictional force proportional to velocity.
Besides the use of fractional derivatives, Bateman [11] suggested
two methods based on the idea that a Lagrangian could lead to
multiple equations. In Refs. [12, 13], Schr\"{o}dinger's equation
was modified directly. For example, a nonlinear term proportional
to $\ln(\psi/\psi^*)$ can be added, and these types of
modifications provide quantum results corresponding to classical
friction. A standard device for dealing with dissipation is the
Rayleigh dissipation function [14], which can be used when
frictional forces are proportional to velocity.

In this paper, a new quantum wave equation is proposed by
application of the Feynman's path-integral. The wave equation can
describe the particle suffering the non-conservative force which
is proportional to velocity.

\section * {2. The Lagrange function for conservative and non-conservative
system}

\hspace{0.3in}The general Lagrange's equations are
\begin{equation}
\frac{d}{dt}\frac{\partial T}{\partial \dot{q}_{j}}-\frac{\partial
T}{\partial q_{j}}=Q_{j} \hspace{0.5in}
(j=1,2,\cdot\cdot\cdot,s),
\end{equation}
where the kinetic energy $T$ and the generalized forces $Q_{j}$
are, in general functions of all the generalized coordinates
$q_{j}$ and generalized velocities $\dot{q}_{j}$and of $t$. The
$Q_{j}$ is defined as
\begin{equation}
Q_{j}= \sum_{i=1}^{N}\vec F_{i}\cdot \frac{\partial
\vec{r_{i}}}{\partial q_{j}},
\end{equation}
where $\vec{r}_{i}=\vec{r}_{i}(q_{1},q_{2},\cdot\cdot\cdot
q_{s},t)$ $(i=1,2,\cdot\cdot\cdot N)$ and $\vec F_{i}$ is the
external force acting on a particle. If all the $\vec F_{i}$ are
conservative force,  and then $\vec F_{i}=-\nabla_{i}V$, we can
write
\begin{equation}
Q_{j}= \sum_{i}\vec F_{i}\cdot \frac{\partial
\vec{r_{i}}}{\partial
q_{j}}=-\sum_{i}\nabla_{i}V\cdot\frac{\partial
\vec{r_{i}}}{\partial q_{j}}=-\frac{\partial V}{\partial q_{j}},
\end{equation}
where the $V$ is a scalar potential function, and so Eq. (1)
becomes
\begin{equation}
\frac{d}{dt}\frac{\partial T}{\partial \dot{q}_{j}}-\frac{\partial
T}{\partial q_{j}}+\frac{\partial V}{\partial q_{j}}= 0,
\end{equation}
since $V$ is a function of the $\vec r_{i}$ and therefore only of
the $q_{j}$ and not the $\dot{q_{j}}$ , we have $\frac{\partial
V}{\partial \dot{q_{j}}}=0$. Let us now define the Lagrangian
function $L$ and action $S$ of the system as
\begin{equation}
L=T-V,
\end{equation}
and
\begin{equation}
S=\int^{t_{2}}_{t_{1}}Ldt,
\end{equation}
and then in terms of this function the equations of motion become
\begin{equation}
\frac{d}{dt}\frac{\partial L}{\partial  \dot q_{j}}-\frac{\partial
L}{\partial q_{j}}= 0 \hspace{0.3in}(j=1,2, \cdot\cdot\cdot,s).
\end{equation}
These are called Lagrange's equations. If the $\vec
F_{i}(i=1,2,\cdot\cdot\cdot N)$ include conservative force and
non-conservative force, we can have
\begin{eqnarray}
Q_{j}=Q_{j}^{(1)}+Q_{j}^{(2)},
\end{eqnarray}
and
\begin{eqnarray}
Q_{j}^{(1)}=-\frac{\partial V}{\partial q_{j}},
\end{eqnarray}
where the $Q_{j}^{(1)}$ is conservative generalized force, and $
Q_{j}^{(2)}$ is non-conservative generalized force. The Lagrange's
equations become
\begin{eqnarray}
\frac{d}{dt}\frac{\partial L}{\partial \dot q_{j}}-\frac{\partial
L}{\partial
q_{j}}=Q_{j}^{(2)}\hspace{0.3in}(j=1,2,\cdot\cdot\cdot,s).
\end{eqnarray}
For a conservative system, we can define the Lagrangian function
$L$ and action S as Eq.(5) and (6). For a generalized system,
i.e.,including conservative and non-conservative force, we can
define the general Lagrangian function $L$ as:
\begin{eqnarray}
L=T+\int^{A}_{P_{0}}\vec{F}_{c}\cdot d \vec
r-\int^{A}_{P_{0}}\vec{F}_{nc}\cdot d \vec r,
\end{eqnarray}
where the $\vec{F}_{c}$ is conservative force and the
$\vec{F}_{nc}$ is non-conservative force, the point $P_{0}$ is the
reference point, and the $A$ is the position point for a particle
in space. If the system is conservative, the Eq. (11) becomes
\begin{eqnarray}
L&=&T+\int^{A}_{P_{0}}\vec F\cdot d \vec r \nonumber\\&=&T-V_A
\end{eqnarray}
where $V_A$ is potential energy of the system. In this case, the
Eq. (11) becomes Eq. (5), and then Eq. (11) can be written as:
\begin{eqnarray}
L=T-V-\int^{A}_{P_0}\vec F_{nc}\cdot d \vec r.
\end{eqnarray}

\section * {3. The quantum wave equation for non-conservative system}

\hspace{0.3in}In the following, we will deduce the quantum wave
equation for the system that there are both conservative and
non-conservative force with the approach of path integral. The
path-integral formula is
\begin{equation}
\Psi(\vec{r},t')=\int
\int\exp[\frac{i}{\hbar}\int^{t'}_{t}L(\dot{\vec{r}}(\tau),{\vec{r}}(\tau),\tau)d\tau]
D[\vec{r}(t)]\Psi(\overrightarrow{r}',t)d\overrightarrow{r}',
\end{equation}
which gives the wave function at a time $t'$ in terms of the wave
function at a time $t$. In order to obtain the differential
equation, we apply this relationship in the special case that the
time $t'$ differs only by an infinitesimal interval $\varepsilon$
from $t$. For a short interval $\varepsilon$ the action is
approximately $\varepsilon$ times the Lagrangian for this
interval, we have
\begin{equation}
\Psi(\vec{r},t+\varepsilon)=\int\frac{d\overrightarrow{r}'}{A^{3}}\exp[\frac{i\varepsilon}{\hbar}
L(\frac{\vec{r}-\overrightarrow{r}'}{\varepsilon},\frac{\vec{r}+\overrightarrow{r}'}{2},\frac{t'+t}{2})]\Psi(\overrightarrow{r}',t),
\end{equation}
where $A$ is a normalization constant. Substituting Eq. (13) into
(15), one can obtain
\begin{equation}
\Psi(\vec{r},t+\varepsilon)=\int\frac{d\overrightarrow{r}'}{A^{3}}\exp[\frac{i\varepsilon}{\hbar}(\frac{m}{2}(\frac{\vec{r}-\overrightarrow{r}'}{\varepsilon})^{2}-
V(\frac{\vec{r}+\overrightarrow{r}'}{2},\frac{t'+t}{2})-\int^{\vec{r}}_{\overrightarrow{r}'}\vec{F}\cdot
d\overrightarrow{r}'')]\Psi(\overrightarrow{r}',t).
\end{equation}
In macroscopic field, the frictional force and adhere force are
non-conservative force, and the non-conservative force $\vec{F}$
is directly proportional to velocity $\vec{v}$, their directions
are opposite, i.e., $\vec{F}=-k\vec{v}$. In microscopic field, a
particle can also suffer the action of non-conservative force. The
nonconservative quantum processes are common too, since there is
dissipation in every nonequilibrium or fluctuating process,
including tunnelling [15], electromagnetic cavity radiation [16,
17], masers and parametric amplification [17], Brownian motion
[12, 18], inelastic scattering [13, 19], squeezed states of
quantum optics [20], and electrical resistance or Ohmic friction
[21]. In the experiment of Bose-Einstein condensates, the atomic
$Rb^{87}$, $Na^{23}$ and $Li^{7}$ can be cooled in laser field,
since they get the non-conservative force from the photons, and
the force $\vec{F}=-k \vec{v}$. Substituting $\vec{F}=-k\vec{v}$
into Eq. (16), we have
\begin{eqnarray}
\Psi(\vec{r},t+\varepsilon)&=&\int\frac{d\overrightarrow{r}'}{A^{3}}\exp[\frac{i\varepsilon}{\hbar}
(\frac{m}{2}(\frac{\vec{r}-\overrightarrow{r}'}{\varepsilon})^{2}-V(\frac{\vec{r}
+\overrightarrow{r}'}{2},\frac{t'+t}{2}) \nonumber\\&& +
k\int^{\vec{r}}_{\overrightarrow{r}'}(\frac{\vec{r}-\overrightarrow{r}'}{\varepsilon})\cdot
d\overrightarrow{r}^{''})]\Psi(\overrightarrow{r}',t)\nonumber\\
&=&\int\frac{d
\overrightarrow{r}'}{A^{3}}exp[\frac{i}{\hbar}\frac{m}{2}\frac{(\vec{r}-\overrightarrow{r}')^{2}}{\varepsilon}]\cdot
exp[-\frac{i
\varepsilon}{\hbar}V(\frac{\vec{r}+\overrightarrow{r}'}{2},\frac{t+t'}{2})]\nonumber\\
&&\cdot exp[\frac{i \varepsilon}{\hbar} k
\int^{\vec{r}}_{\overrightarrow{r}'}\frac{\vec{r}-\overrightarrow{r}'}{\varepsilon}\cdot
d\overrightarrow{r}'']\cdot \psi(\overrightarrow{r}',t).
\end{eqnarray}
The quantity
$\frac{({\overrightarrow{r}-\overrightarrow{r}'})^{2}}{\varepsilon}$
appears in the exponent of the first factor. It is clear that if
$\overrightarrow{r}'$ is appreciably different from $\vec{r}$,
this quantity is very large and the exponential consequently
oscillates very rapidly as $\overrightarrow{r}'$ varies, when this
factor oscillates rapidly, the integral over $\overrightarrow{r}'$
gives a very small value. Only if $\overrightarrow{r}'$ is near
$\vec{r}$ do we get important contributions. For this reason we
make the substitution $\overrightarrow{r}'=\vec{r}+\vec{\eta}$
with the expectation that appreciable contribution to the integral
will occur only for small $\vec{\eta}$, we obtain
\begin{eqnarray}
\Psi(\vec{r},t+\varepsilon)&=&\int\frac{d\vec{\eta}}{A^{3}}\exp[\frac{i}{\hbar}
\frac{m}{2} \frac{\vec{\eta}^{2}}{\varepsilon}]\cdot exp[-\frac{i
\varepsilon}{\hbar}V(\vec{r}+\frac{\vec{\eta}}{2},t+\frac{\varepsilon}
{2})] \nonumber\\
&& \cdot exp[\frac{i
\varepsilon}{\hbar}k\int^{\vec{r}}_{\overrightarrow{r}'}\frac{-\vec{\eta}}{\varepsilon}\cdot
d\overrightarrow{r}'')]\Psi(\vec{r}+\vec{\eta},t)
\end{eqnarray}
Now we have
\begin{eqnarray}
\int^{\vec{r}}_{\overrightarrow{r}'}\vec{\eta}\cdot
d\overrightarrow{r}''=\int^{\vec{r}}_{\overrightarrow{r}'}|\vec{\eta}|
|d\overrightarrow{r}''|\cos\theta=|\vec{\eta}|\int^{\vec{r}}_{\overrightarrow{r}'}
|d\overrightarrow{r}''|\cos\theta=-|\vec{\eta}|^{2}
\end{eqnarray}
so that
\begin{eqnarray}
k\int^{\vec{r}}_{\overrightarrow{r}'}\frac{-\vec{\eta}}{\varepsilon}
\cdot d\overrightarrow{r}''=\frac{k}{\varepsilon}
|\vec{\eta}|^{2}=\frac{k}{\varepsilon}\vec{\eta}^{2}
\end{eqnarray}
substituting Eq. (20) into (18), we have
\begin{eqnarray}
\Psi(\vec{r},t+\varepsilon)
=\int\frac{d\vec{\eta}}{A^{3}}e^{\frac{im\vec{\eta}^{2}}{2\hbar\varepsilon}}
e^{-\frac{i\varepsilon}{\hbar}V(\vec{r}+\frac{\vec{\eta}}{2},t+\frac{\varepsilon}{2})}
e^{\frac{i}{\hbar}k\vec{\eta}^{2}} \Psi(\vec{r}+\vec{\eta},t)
\end{eqnarray}
The phase of the first exponential changes by the order of 1
radian when $|\vec{\eta}|$ is of the order
$\sqrt{\frac{2\hbar\varepsilon}{m}}$, so that most of the integral
is contributed by values of $|\vec{\eta}|$ in this order. We may
expand $\Psi$ in a power series, we need only keep terms of order
$\varepsilon$. This implies keeping second-order terms in
$\eta_{x}$, $\eta_{y}$ and $\eta_{z}$. Expanding the left-hand
side to first order in $\varepsilon$ and the right-hand side to
first order in $\varepsilon$ and second order in $\eta_{x}$,
$\eta_{y}$ and $\eta_{z}$, we have
\begin{equation}
e^{-\frac{i\varepsilon}{\hbar}V(\vec{r}+\frac{\vec{\eta}}{2},t+\frac{\varepsilon}{2})}=1-\frac{i\varepsilon}{\hbar}V(\vec{r},t)
\end{equation}
\begin{equation}
e^{\frac{i}{\hbar}k\vec{\eta}^{2}}=1+\frac{i}{\hbar}k\vec{\eta}^{2}
\end{equation}

\begin{equation}
\Psi(\vec{r}+\vec{\eta},t)=\Psi(\vec{r},t)+ \vec{\eta}\cdot
\frac{\partial\Psi(\vec{r},t)}{\partial\vec{r}}+
 \frac{1}{2}(\eta^{2}_{x}\frac{\partial^{2}\Psi}{\partial x^{2}}+
 \eta^{2}_{y}\frac{\partial^{2}\Psi}{\partial y^{2}}+\eta^{2}_{z}\frac{\partial^{2}
 \Psi}{\partial z^{2}})
\end{equation}
and
\begin{eqnarray}
\Psi(\vec{r},t)+\varepsilon\frac{\partial\Psi(\vec{r},t)}{\partial
t}&=&\int\frac{d\vec{\eta}}{A^{3}}
e^{\frac{im\vec{\eta}^{2}}{2\hbar\varepsilon}}
(1-\frac{i\varepsilon}{\hbar}V(\vec{r},t))
(1+\frac{i}{\hbar}k\vec{\eta}^{2})  [\Psi(\vec{r},t)+
\vec{\eta}\cdot \frac{\partial\Psi(\vec{r},t)}{\partial\vec{r}}+
\nonumber\\&&
 \frac{1}{2}(\eta^{2}_{x}\frac{\partial^{2}\Psi}{\partial x^{2}}+
 \eta^{2}_{y}\frac{\partial^{2}\Psi}{\partial y^{2}}+\eta^{2}_{z}\frac{\partial^{2}
 \Psi}{\partial z^{2}})]
\nonumber\\
 &=&\int\frac{d\vec{\eta}}{A^{3}}e^{\frac{im\vec{\eta}^{2}}{2\hbar\varepsilon}}
(1+\frac{i}{\hbar}k\vec{\eta}^{2}-\frac{i\varepsilon}{\hbar}V
(\vec{r},t)+\frac{\varepsilon}{\hbar^{2}}k
V(\vec{r},t)\vec{\eta}^{2})
\nonumber\\
&&[\Psi(\vec{r},t)+\vec{\eta}\cdot
\nabla\Psi(\vec{r},t)+\frac{1}{2}(\eta^{2}_{x}\frac{\partial^{2}\Psi}{\partial
x^{2}}+\eta^{2}_{y}\frac{\partial^{2}\Psi}{\partial
y^{2}}+\eta^{2}_{z}\frac{\partial^{2}\Psi}{\partial z^{2}})]
\end{eqnarray}
In order to evaluate the right-hand side of Eq. (25), we shall
have to use four integrals
\begin{equation}
\int^{\infty}_{-\infty}d\eta_{x}e^{\frac{i m
\eta^{2}_{x}}{2\hbar\varepsilon}}\equiv
A=(\frac{i2\pi\hbar\varepsilon}{m})^{\frac{1}{2}},
\end{equation}
\begin{equation}
\int^{\infty}_{-\infty}d\eta_{x}
\eta_{x}e^{\frac{im\eta^{2}_{x}}{2\hbar\varepsilon}}=0,
\end{equation}
\begin{equation}
\int^{\infty}_{-\infty}d\vec{\eta}e^{\frac{im\vec{\eta}^{2}}{2\hbar\varepsilon}}=
A^{3}=(\frac{i2\pi\hbar\varepsilon}{m})^{\frac{3}{2}},
\end{equation}
and
\begin{equation}
\int^{\infty}_{-\infty}d\eta_{x}
\eta^{2}_{x}e^{\frac{im\eta^{2}_{x}}{2\hbar\varepsilon}}=\frac{i\hbar\varepsilon}{m}(\frac{i2\pi\hbar\varepsilon}{m})^{\frac{1}{2}}
\end{equation}
In Eq. (25), we can easily find that the integrals of the terms
$\frac{\varepsilon}{\hbar^{2}}k
V(\vec{r},t)\overrightarrow{\eta}^{2}$ and
$\vec{\eta}\cdot\nabla\psi(\vec{r}, t)$ are either zero or
$O(\varepsilon^{2})$ from Eqs. (26)-(29), and they can be
neglected in Eq. (25). The Eq. (25) becomes
\begin{eqnarray}
\Psi(\vec{r},t)+\varepsilon\frac{\partial\Psi(\vec{r},t)}{\partial
t}&=&\int\frac{d\vec{\eta}}{A^{3}}e^{\frac{im\vec{\eta}^{2}}{2\hbar\varepsilon}}
[1+\frac{i}{\hbar}k\vec{\eta}^{2}-\frac{i\varepsilon}{\hbar}V(\vec{r},t)]
[\Psi(\vec{r},t)
\nonumber\\
&& +\frac{1}{2}(\eta^{2}_{x}\frac{\partial^{2}\Psi}{\partial
x^{2}}+\eta^{2}_{y}\frac{\partial^{2}\Psi}{\partial
y^{2}}+\eta^{2}_{z}\frac{\partial^{2}\Psi}{\partial z^{2}})]
\nonumber\\
&=&\int\frac{d\vec{\eta}}{A^{3}}e^{\frac{im\vec{\eta}^{2}}{2\hbar\varepsilon}}(1-\frac{i\varepsilon}{\hbar}V(\vec{r},t))
(\Psi(\vec{r},t)
\nonumber\\
&& +\frac{1}{2}(\eta^{2}_{x}\frac{\partial^{2}\Psi}{\partial
x^{2}}+\eta^{2}_{y}\frac{\partial^{2}\Psi}{\partial
y^{2}}+\eta^{2}_{z}\frac{\partial^{2}\Psi}{\partial
z^{2}}))\nonumber\\
&&+\frac{i}{\hbar}k\int\frac{d\vec{\eta}}{A^{3}}e^{\frac{im\eta^{2}}{2\hbar\varepsilon}}
\vec{\eta}^{2}[\Psi(\vec{r},t)
\nonumber\\
&& +\frac{1}{2}(\eta^{2}_{x}\frac{\partial^{2}\Psi}{\partial
x^{2}}+\eta^{2}_{y}\frac{\partial^{2}\Psi}{\partial
y^{2}}+\eta^{2}_{z}\frac{\partial^{2}\Psi}{\partial z^{2}})].
\end{eqnarray}
In Eq. (30), the first term is
\begin{eqnarray}
&&\int\frac{d\vec{\eta}}{A^{3}}e^{\frac{im\vec{\eta}^{2}}{2\hbar\varepsilon}}(1-\frac{i\varepsilon}{\hbar}V(\vec{r},t))
\Psi(\vec{r},t)
\nonumber\\&=&(1-\frac{i\varepsilon}{\hbar}V(\vec{r},t)) \Psi(\vec{r},t) \int\frac{d\vec{\eta}}
{A^{3}}e^{\frac{im\vec{\eta}^{2}}{2\hbar\varepsilon}}\nonumber\\
&=&(1-\frac{i\varepsilon}{\hbar}V(\vec{r},t)) \Psi(\vec{r},t),
\end{eqnarray}
and the second term is
\begin{eqnarray}
&&\frac{1}{2}\int\frac{d\vec{\eta}}{A^{3}}e^{\frac{im\vec{\eta}^{2}}{2\hbar\varepsilon}}
(1-\frac{i\varepsilon}{\hbar}V(\vec{r},t))(\eta^{2}_{x}\frac{\partial^{2}\Psi}{\partial
x^{2}}+\eta^{2}_{y}\frac{\partial^{2}\Psi}{\partial
y^{2}}+\eta^{2}_{z}\frac{\partial^{2}\Psi}{\partial z^{2}})
\nonumber\\&=&\frac{1}{2}\int\frac{d\vec{\eta}}{A^{3}}e^{\frac{im\vec{\eta}^{2}}{2\hbar\varepsilon}}
\eta^{2}_{x}\frac{\partial^{2}\Psi}{\partial
x^{2}}+\frac{1}{2}\int\frac{d\vec{\eta}}{A^{3}}e^{\frac{im\vec{\eta}^{2}}{2\hbar\varepsilon}}
\eta^{2}_{y}\frac{\partial^{2}\Psi}{\partial
y^{2}}+\frac{1}{2}\int\frac{d\vec{\eta}}{A^{3}}e^{\frac{im\vec{\eta}^{2}}{2\hbar\varepsilon}}
\eta^{2}_{z}\frac{\partial^{2}\Psi}{\partial z^{2}}.
\end{eqnarray}
In Eq. (32), we can easily find that the integral of the term
$\frac{i\varepsilon}{\hbar}V(\vec{r},t)$ is $O(\varepsilon^{2})$,
which can be neglected. In Eq. (32), the first term is
\begin{eqnarray}
\frac{1}{2}\int\frac{d\vec{\eta}}{A^{3}}e^{\frac{im\vec{\eta}^{2}}{2\hbar\varepsilon}}
\eta^{2}_{x}\frac{\partial^{2}\Psi}{\partial
x^{2}}=\frac{1}{2}\frac{\partial^{2}\Psi}{\partial
x^{2}}\int\frac{d\eta_{x}}{A}\eta^{2}_{x}e^{\frac{im\eta^{2}_{x}}{2\hbar\varepsilon}}
\int\frac{d\eta_{y}}{A}e^{\frac{im\eta^{2}_{y}}{2\hbar\varepsilon}}
 \int\frac{d\eta_{z}}{A}e^{\frac{im\eta^{2}_{z}}{2\hbar\varepsilon}}=\frac{1}{2}\frac{\partial^{2}\Psi}{\partial
x^{2}} \frac{i\hbar\varepsilon}{m},
\end{eqnarray}
and similarly, the second and third terms are
\begin{eqnarray}
&&\frac{1}{2}\int\frac{d\vec{\eta}}{A^{3}}e^{\frac{im\vec{\eta}^{2}}{2\hbar\varepsilon}}
\eta^{2}_{y}\frac{\partial^{2}\Psi}{\partial
y^{2}}=\frac{1}{2}\frac{\partial^{2}\Psi}{\partial
y^{2}} \frac{i\hbar\varepsilon}{m},\nonumber\\
&&\frac{1}{2}\int\frac{d\vec{\eta}}{A^{3}}e^{\frac{im\vec{\eta}^{2}}{2\hbar\varepsilon}}
\eta^{2}_{z}\frac{\partial^{2}\Psi}{\partial
z^{2}}=\frac{1}{2}\frac{\partial^{2}\Psi}{\partial z^{2}}
\frac{i\hbar\varepsilon}{m},
\end{eqnarray}
and so Eq. (32) becomes
\begin{eqnarray}
\frac{1}{2}\int\frac{d\vec{\eta}}{A^{3}}e^{\frac{im\vec{\eta}^{2}}{2\hbar\varepsilon}}
(\eta^{2}_{x}\frac{\partial^{2}\Psi}{\partial
x^{2}}+\eta^{2}_{y}\frac{\partial^{2}\Psi}{\partial
y^{2}}+\eta^{2}_{z}\frac{\partial^{2}\Psi}{\partial
z^{2}})=\frac{1}{2}\frac{i\hbar\varepsilon}{m}(\frac{\partial^{2}\Psi}{\partial
x^{2}}+\frac{\partial^{2}\Psi}{\partial
y^{2}}+\frac{\partial^{2}\Psi}{\partial
z^{2}})=\frac{i\hbar}{2m}\varepsilon\nabla^{2}\Psi.
\end{eqnarray}
In Eq. (30), the third term is
\begin{eqnarray}
&&\frac{i}{\hbar}k\int\frac{d\vec{\eta}}{A^{3}}e^{\frac{im\vec{\eta}^{2}}{2\hbar\varepsilon}}\vec{\eta}^{2}
\Psi(\vec{r},t)
\nonumber\\&=&\frac{i}{\hbar}k\Psi(\vec{r},t)[\int\frac{d\vec{\eta}}{A^{3}}e^{\frac{im\vec{\eta}^{2}}{2\hbar\varepsilon}}\eta^{2}_{x}
+\int\frac{d\vec{\eta}}{A^{3}}e^{\frac{im\vec{\eta}^{2}}{2\hbar\varepsilon}}\eta^{2}_{y}
+\int\frac{d\vec{\eta}}{A^{3}}e^{\frac{im\vec{\eta}^{2}}{2\hbar\varepsilon}}\eta^{2}_{z}].
\end{eqnarray}
In Eq. (36), the first term is
\begin{eqnarray}
\int\frac{d\vec{\eta}}{A^{3}}e^{\frac{im\vec{\eta}^{2}}{2\hbar\varepsilon}}\eta^{2}_{x}
=\int\frac{d\eta_{x}}{A}\eta^{2}_{x}e^{\frac{im\eta^{2}_{x}}{2\hbar\varepsilon}}
 \int\frac{d\eta_{y}}{A}e^{\frac{im\eta^{2}_{y}}{2\hbar\varepsilon}}
 \int\frac{d\eta_{z}}{A}e^{\frac{im\eta^{2}_{z}}{2\hbar\varepsilon}}
=\frac{i\hbar\varepsilon}{m},
\end{eqnarray}
and similarly, the second and third terms are
\begin{eqnarray}
&&\int\frac{d\vec{\eta}}{A^{3}}e^{\frac{im\vec{\eta}^{2}}{2\hbar\varepsilon}}
\eta^{2}_{y}=\frac{i\hbar\varepsilon}{m},\nonumber\\
&&\int\frac{d\vec{\eta}}{A^{3}}e^{\frac{im\vec{\eta}^{2}}{2\hbar\varepsilon}}
\eta^{2}_{z}=\frac{i\hbar\varepsilon}{m},
\end{eqnarray}
and so Eq. (36) becomes
\begin{eqnarray}
\frac{i}{\hbar}k\int\frac{d\vec{\eta}}{A^{3}}e^{\frac{im\vec{\eta}^{2}}{2\hbar\varepsilon}}\vec{\eta}^{2}
\Psi(\vec{r},t) =\frac{i}{\hbar}k\Psi(\vec{r},t)
\frac{3i\hbar\varepsilon}{m} =-\frac{3\varepsilon
k}{m}\Psi(\vec{r},t).
\end{eqnarray}
In Eq. (30), the fourth term is
\begin{eqnarray}
&&\frac{i}{\hbar}k\int\frac{d\vec{\eta}}{A^{3}}e^{\frac{im\vec{\eta}^{2}}{2\hbar\varepsilon}}
\vec{\eta}^{2}
\frac{1}{2}(\eta^{2}_{x}\frac{\partial^{2}\Psi}{\partial
x^{2}}+\eta^{2}_{y}\frac{\partial^{2}\Psi}{\partial
y^{2}}+\eta^{2}_{z}\frac{\partial^{2}\Psi}{\partial z^{2}})
\nonumber\\&=&\frac{i}{2\hbar}k\int\frac{d\vec{\eta}}{A^{3}}e^{\frac{im\vec{\eta}^{2}}{2\hbar\varepsilon}}\vec{\eta}^{2}
 \eta^{2}_{x}\frac{\partial^{2}\Psi}{\partial x^{2}}+\frac{i}{2\hbar}k\int\frac{d\vec{\eta}}{A^{3}}e^{\frac{im\vec{\eta}^{2}}
 {2\hbar\varepsilon}}\vec{\eta}^{2}
 \eta^{2}_{y}\frac{\partial^{2}\Psi}{\partial y^{2}}\nonumber\\
&&+\frac{i}{2\hbar}k\int\frac{d\vec{\eta}}{A^{3}}e^{\frac{im\vec{\eta}^{2}}{2\hbar\varepsilon}}\vec{\eta}^{2}
 \eta^{2}_{z}\frac{\partial^{2}\Psi}{\partial z^{2}}.
\end{eqnarray}
In Eq. (40), the first term is
\begin{eqnarray}
&&\frac{i}{2\hbar}k\int\frac{d\vec{\eta}}{A^{3}}e^{\frac{im\vec{\eta}^{2}}{2\hbar\varepsilon}}\vec{\eta}^{2}
 \eta^{2}_{x}\frac{\partial^{2}\Psi}{\partial x^{2}}
\nonumber\\
&=&\frac{i}{2\hbar}k\frac{\partial^{2}\Psi}{\partial
x^{2}}[\int\frac{d\vec{\eta}}{A^{3}}e^{\frac{im\vec{\eta}^{2}}{2\hbar\varepsilon}}(\eta^{4}_{x}+\eta^{2}_{x}\eta^{2}_{y}+\eta^{2}_{x}\eta^{2}_{z})]
\nonumber\\&=&\frac{i}{2\hbar}k\frac{\partial^{2}\Psi}{\partial
x^{2}}[\int\frac{d{\eta_{x}}}{A}\eta^{4}_{x}e^{\frac{im{\eta}^{2}_{x}}{2\hbar\varepsilon}}
\int\frac{d{\eta_{y}}}{A}e^{\frac{im{\eta}^{2}_{y}}{2\hbar\varepsilon}}
\int\frac{d{\eta_{z}}}{A}e^{\frac{im{\eta}^{2}_{z}}{2\hbar\varepsilon}}\nonumber\\&&+
\int\frac{d{\eta_{x}}}{A}\eta^{2}_{x}e^{\frac{im{\eta}^{2}_{x}}{2\hbar\varepsilon}}
\int\frac{d{\eta_{y}}}{A}\eta^{2}_{y}e^{\frac{im{\eta}^{2}_{y}}{2\hbar\varepsilon}}
\int\frac{d{\eta_{z}}}{A}e^{\frac{im{\eta}^{2}_{z}}{2\hbar\varepsilon}}\nonumber\\&&+
\int\frac{d{\eta_{x}}}{A}\eta^{2}_{x}e^{\frac{im{\eta}^{2}_{x}}{2\hbar\varepsilon}}
\int\frac{d{\eta_{y}}}{A}e^{\frac{im{\eta}^{2}_{y}}{2\hbar\varepsilon}}
\int\frac{d{\eta_{z}}}{A}\eta^{2}_{z}e^{\frac{im{\eta}^{2}_{z}}{2\hbar\varepsilon}}].
\end{eqnarray}
From Eq. (29), we can obtain the integral formula
\begin{eqnarray}
\int\frac{d{\eta_{x}}}{A}\eta^{4}_{x}e^{\frac{im{\eta}^{2}_{x}}{2\hbar\varepsilon}}=3(\frac{i\hbar\varepsilon}{m})^{2}.
\end{eqnarray}
From Eq. (26), (29) and (42), Eq. (41) becomes
\begin{eqnarray}
&=&\frac{i}{2\hbar}k\frac{\partial^{2}\Psi}{\partial
x^{2}}[3(\frac{i\hbar\varepsilon}{m})^{2}+(\frac{i\hbar\varepsilon}{m})^{2}+(\frac{i\hbar\varepsilon}{m})^{2}]
\nonumber\\&=&5(\frac{i\hbar\varepsilon}{m})^{2}
\frac{i}{2\hbar}k\frac{\partial^{2}\Psi}{\partial x^{2}}.
\end{eqnarray}
The Eq. (43) is directly proportional to $\varepsilon^{2}$.
Obviously, in Eq. (40), the second and third terms
 are also directly proportional to $\varepsilon^{2}$, and so the
 contribution of Eq. (40) can be neglected.

Substituting Eq. (31), (35) and (39) into (30), we can obtain
\begin{equation}
\Psi(\vec{r},t)+\varepsilon\frac{\partial\Psi(\vec{r},t)}{\partial
t}=(1-\frac{i\varepsilon}{\hbar}V(r,t))\Psi(\vec{r},t)+\frac{i\hbar}{2m}\varepsilon
\nabla^{2} \Psi -\frac{3 \varepsilon k}{m} \Psi(\vec{r},t).
\end{equation}
Equating the coefficient of powers of $\varepsilon$, we have
\begin{eqnarray}
\frac{\partial\Psi(\vec{r},t)}{\partial
t}=-\frac{i}{\hbar}V(r,t)\Psi(\vec{r},t)+\frac{i\hbar}{2m}\nabla^{2}\Psi
-\frac{3k}{m}\Psi(\vec{r},t),
\end{eqnarray}
multiplied by the coefficient of $i\hbar$, we have
\begin{eqnarray}
i\hbar\frac{\partial\Psi(\vec{r},t)}{\partial
t}&=&(-\frac{\hbar^{2}}{2m}\nabla^{2}+V
-i\hbar\frac{3k}{m})\Psi(\vec{r},t).
\end{eqnarray}
The Eq. (46) is a new quantum wave equation, which is suitable for
the non-conservative force $\vec{F}=-k\vec{v}$. From the equation,
we can study the non-conservative system.

\section * {4. Conclusion}

\hspace{0.3in}We know Schr\"{o}dinger's equation is quantum wave
equation, which is only suitable for a conservative system. For a
non-conservative system, we need a new quantum wave equation. In
this paper, we apply the approach of path integral to obtain the
general quantum wave equation, which is suitable for a
non-conservative system. We think the new quantum wave equation
should be used widely in the future.


\newpage
\begin{thebibliography}{99}
\bibitem{s1}
R. P. Feynman, Rev. Mod. Phys., {\bf 20}, (1948) 367.
\bibitem{s2}
R. P. Feynman and A. R. Hibbs, Quantum mechanics and path
integrals, McGraw-Hill Book Co.New York 1965.
\bibitem{s3}
R. P. Feynman, Phys. Rev, {\bf 80}, (1950) 440.
\bibitem{s4}
P. Jordan, Z. Phys. {\bf 80} (1933) 285; P. Jordan and E. P.
Wigner, Ann. Math. {\bf 35} (1934) 29.
\bibitem{s5}
D. Finkelstein et al., Notes on Quaternion Quantum Mechanics
(CERN, Report 59-7), in C. Hoker, ed. Logico-Algebraic-Approach to
Quantum Mechanics $\coprod$ (Reidel, Dordrecht, 1979).
\bibitem{s6}
D. Finkelstein et al., J. Math. Phys. {\bf 3} (1962) 207; ibid.
{\bf 4} (1963) 788
\bibitem{s7}
S. De Leo and K. Abdel-Khalek, Prog. Theor. Phys. {\bf 96} (1996)
823.
\bibitem{s8}
D. Minic and C. -H. Tze, Phys. Lett. B {\bf 581} (2004) 111.
\bibitem{s9}
F. Riewe, Phys. Rev. E {\bf 53}, (1996) 1890.
\bibitem{s10}
F. Riewe, Phys. Rev. E {\bf 55}, (1997) 3581.
\bibitem{s11}
H. Bateman, Phys. Rev. {\bf 38}, (1931) 815.
\bibitem{s12}
M. Razavy. Can. J. Phys. {\bf 56}, (1978) 311.
\bibitem{s13}
R. W. Hasse, J. Math. Phys. {\bf 16}, (1975) 2005.
\bibitem{s14}
H. Goldstein, Classical Mechanics (Addison-Wesley, Reading, MA,
1950) pp. 21-22.
\bibitem{s15}
A. O. Caldeira and A. J. Leggett, Phys. Rev. Lett., {\bf 46},
(1981) 211; Ann. Phys. ~N. Y. {\bf 149}, (1983) 374; J. Ankerhold,
H. Grabert, and G.-L. Ingold, Phys. Rev. E, {\bf 52}, (1995) 4267.
\bibitem{s16}
I. R. Senitzky, Phys. Rev. {\bf 119}, (1960) 670 .
\bibitem{s17}
K. W. H. Stevens, Proc. Phys. Soc. London {\bf 72}, (1958)1027.
\bibitem{s18}
G. W. Ford, M. Kac, and P. Mazur, J. Math. Phys. {\bf 6}, (1965)
504; P. Ullersma, Physica (Utrecht) {\bf 32}, (1966) 27.
\bibitem{s19}
T. J. Krieger, Phys. Rev. {\bf 121}, (1961) 1388.
\bibitem{s20} E. Celeghini, M.
Rasetti, M. Tarlini, and G. Vitiello, Mod. Phys. Lett. B {\bf 3},
(1989) 1213.
\bibitem{s21} H. B. Callen and T. A. Welton, Phys. Rev. {\bf 83}, (1951) 34;
L. H. Yu and C.-P. Sun, Phys. Rev. A {\bf 49}, (1994) 592.


\end{thebibliography}
\end{document}